\documentclass[twocolumn,showpacs,preprintnumbers,superscriptaddress,amsmath,amssymb]{revtex4-1}

\usepackage{epsfig}
\usepackage{graphicx}% Include figure files
\usepackage{dcolumn}% Align table columns on decimal point
\usepackage{bm}% bold math
\usepackage{times}
\usepackage{xcolor}

\begin{document}

%\preprint{APS/123-QED}

\title{Social contagions on time-varying community
networks}

\author{Mian-Xin Liu}
\affiliation{Web Sciences Center, University of Electronic
Science and Technology of China, Chengdu 610054, China}
\affiliation{Big data research center, University of Electronic Science and Technology of China, Chengdu
611731, China}

\author{Wei Wang} \email{wwzqbx@hotmail.com}
\affiliation{Web Sciences Center, University of Electronic
Science and Technology of China, Chengdu 610054, China}
\affiliation{Big data research center, University of Electronic Science and Technology of China, Chengdu
611731, China}
\author{Ying Liu}
\affiliation{Web Sciences Center, University of Electronic
Science and Technology of China, Chengdu 610054, China}
\affiliation{Big data research center, University of Electronic Science and Technology of China, Chengdu
611731, China}
\affiliation{School of Computer Science, Southwest Petroleum University, Chengdu 610500, China}

\author{Ming Tang} \email{tangminghan007@gmail.com}
\affiliation{Web Sciences Center, University of Electronic
Science and Technology of China, Chengdu 610054, China}
\affiliation{Big data research center, University of Electronic Science and Technology of China, Chengdu
611731, China}

\author{Shi-Min Cai}
\affiliation{Web Sciences Center, University of Electronic
Science and Technology of China, Chengdu 610054, China}
\affiliation{Big data research center, University of Electronic Science and Technology of China, Chengdu
611731, China}

\author{Hai-Feng Zhang}
\affiliation{School of Mathematical Science, Anhui University, Hefei 230039, China}

\date{\today}

\begin{abstract}
Time-varying community structures widely exist in various real-world networks. However, the spreading dynamics on this kind of network has not been fully studied.
To this end, we systematically study the effects of time-varying community structures
on social contagions. We first propose a non-Markovian social contagion model
on time-varying community networks based
on the activity driven network model, in which an individual adopts a behavior if and
only if the accumulated behavioral information it has ever received
reaches a threshold.
Then, we develop a mean-field theory to describe the proposed model. From theoretical
 analyses and numerical simulations, we find that behavior adoption in
the social contagions exhibits a hierarchical feature, i.e., the behavior first quickly spreads in one of the communities, and then outbreaks in the other. Moreover,
under different behavioral information transmission rates, the final behavior adoption proportion in the
whole network versus the community strength shows one of the patterns, which are a monotone increasing pattern,
a non-monotonic changing pattern, and a monotone decreasing pattern. An optimal community
strength maximizing the final behavior adoption can be found in a suitable
range of behavioral information
transmission rate. Finally, for a given average
degree, increasing the number of edges generated by active nodes
is more beneficial to the social contagions
than increasing the average activity potential.

\end{abstract}

\maketitle

\section{Introduction} \label{sec:intro}
The spreading dynamics is one of the hottest research
topics in network science, which has attracted extensive attention from
scholars in physics, mathematics, biology and other fields.
The spreading dynamics aims to reveal the mechanisms in
real spreading processes such as epidemic spreading, information spreading,
behavior contagion and innovation diffusion, and further provides the theoretical support
for forecasting and controlling these processes~\cite{Pastor-Satorras2015}. The spreading dynamics can be divided into
biological spreading and social contagion. The former
focuses on the spreads of disease or virus on networks~\cite{
Pastor-Satorras2001,Newman2002a,Gross2006,Small2007}, while the latter mainly
studies contagions of information and behaviors on networks~\cite{young2011,
Centola2011,Adam2014,Gleeson2016}. The social reinforcement
effect in social contagion is the essential difference between biological spreading and social contagion~\cite{RMP2009}, which contains the idea that adoption behaviors of an individual
often depends on his neighbors' attitudes to the behavior~\cite{watts2004,
watts2005,centola2010}. For an individual, who has two friends having adopted a particular
behavior before a \emph{given time} and whose third friend newly adopts the
behavior, his/her decision to adopt this behavior will take all the three friends
into account.

For social contagions, researches focus on how
social reinforcement effect influences the spreads of behaviors
on static networks. The Markovian linear threshold model is a
classic social contagion model to describe this reinforcement effect~\cite{watts2002}.
In the model, an individual that has not adopted a behavior becomes an adopter only when the number or proportion of
its adopted neighbors exceeds a threshold. Watts found that the final behavior
adoption proportion, following the increase of average degree, first
grows continuously and then decreases discontinuously~\cite{watts2002}.
In fact, an individual's decision to adopt a behavior not only depends on the
\emph{current state} of his/her neighbors, but also considers the behavioral
information he has received. So the social reinforcement
effect based on memory thus becomes an essential characteristic of social
contagions. To describe the memory effect (i.e., non-Markovian
effect), Wang \emph{et al.} proposed a social contagion model based on
non-redundant memory information, and found that the behavior adoption
proportion versus the information transmission rate could exhibit
a continuous growth or behavior as a discontinuous
growth~\cite{wangwei2015a,Wang2016}.
They also found that the individual's limited contact capacity would
reduce the final behavior adoption proportion~\cite{Wangwei2015b}.

The latest empirical studies showed that the connections among
individuals in social networks vary with
time, which can not be described
by the static network. To this end, the
conception of time-varying network (or temporal networks, dynamical networks) was proposed~\cite{HolmePR}. Perra \emph{et al.} proposed an activity-driven network model to describe time-varying networks~\cite{Perra2012,PerraPRL},
which allows for an explicit representation of dynamical
connectivity patterns.
At each time step, every node becomes active or not according to its
active potential. If a node becomes active, it will randomly connect to some nodes
and form an instantaneous network structure. Spreading processes in
activity-driven networks model show striking differences with respect to the
well-known results obtained in quenched and annealed networks. Perra \emph{et al.} found that the outbreak
threshold of SIS model on an activity driven network is greater than
that of the corresponding aggregated network~\cite{Perra2012}. Liu \emph{et al.} found that a disease spreads slower
on activity driven networks than it does on the corresponding aggregated
networks, and the invasion threshold on the former was hundreds of times
greater than that of the latter~\cite{LiuSY2013}. Holme
\emph{et al.} studied the threshold model on the time-varying networks
based on empirical data, and found that time-varying network structures could
enhance the final behavior adoption proportion~\cite{holme2013}.

Community structures exist ubiquitously in real world networks~\cite{Newman2002,
Newman2006}, greatly affecting the spreading dynamics. For example,
Liu \emph{et al.} found that community structures
make the epidemic spread more easily on static networks~\cite{Liu2005}, and Ahn \emph{et al.} found that there is an optimal community strength
which can greatly promote the social contagions~\cite{Ann2014}. Recent empirical
studies showed that community structures also exists on time-varying networks~\cite{HolmePR,Mucha2012}. However, the effects of time-varying
community structures on social contagion are
little studied and full of challenges. On the one hand, the contacts
on time-varying community networks change over time and do not happen continuously. On the other hand, the social reinforcement effects
lead to the non-Markovian characteristic, making
the existing theoretical method on static network difficult
to accurately describe the spreading processes. In this paper,
we systematically study the effects of time-varying community
structures on social contagions. Firstly, we propose a non-Markovian
social contagion model on time-varying community networks.
Then, we develop a mean-field theory to quantify this contagion
process and verify the accuracy of our predictions via extensive
numerical simulations. With analyses and simulations, we
find that behavior adoption exhibits a
hierarchical feature: the behavior first spreads in
one of the communities, and then outbreaks in the other. Moreover, under
different transmission rates, the final behavior adoption proportion
in the whole network versus community strength shows one of the following three patterns, which are an
increasing pattern, a non-monotonic changing pattern, and a monotone decreasing pattern. An optimal
community strength maximizing the final behavior adoption proportion can be
found in a suitable transmission rate range. Moreover, we find
that for a given average degree, adding the edges
generated by active individuals is more beneficial to social contagions than increasing the
average activity potential.

\section{Models} \label{sec:model}
In order to study the effects of time-varying community networks on social contagions, we propose a non-Markovian social contagion
model on activity-driven community network.

\subsection{Activity-driven community network}
\label{2.1}
\begin{figure}
\begin{center}
\epsfig{file=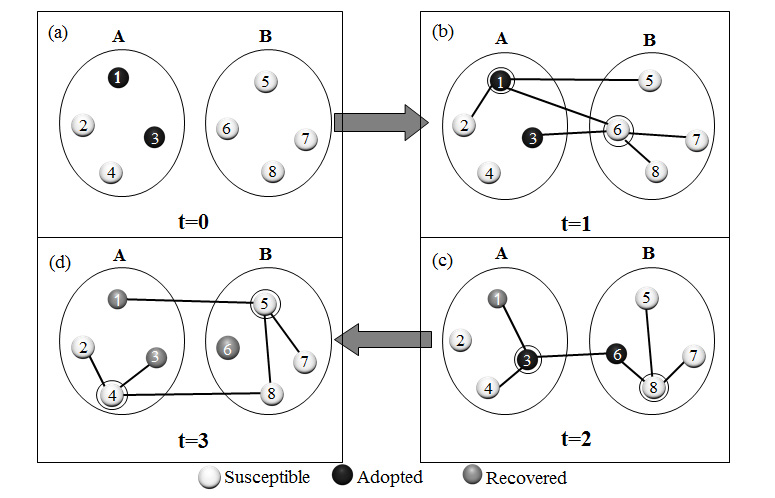,width=1\linewidth}
\caption{(Color online) An illustration of social contagion model on activity-driven
community network. The network is divided into two equal sized communities $A$ and $B$, each of which has 4 nodes.
The circle around node reflects that node is active. (a) At $t=0$, randomly choose
nodes 1 and 3 as seeds on community
$A$, and the remaining nodes are susceptible.
(b) At $t=1$, generate the
instantaneous structure $G_1$, in which
nodes 1 and 6 are activated with probability $a=0.25$ and generate
$m=3$ edges. Every edge connects to nodes in the same community with probability $u=0.6$ and with probability $1-u=0.4$ to the other
community. Adopted nodes 1 and 3 transmit the behavioral information to susceptible neighbors with $\lambda=0.8$. Node 6 receives 2
pieces of information successfully, and reaches the adoption
threshold $\pi=2$, thus it becomes adopted.
Nodes 1 and 3 become recovered
with $\gamma=0.1$. Delete all edges
generated at this time step. (c) At $t=2$, nodes 3 and 8 are active and
form the instantaneous structure $G_2$.
Node 3 transmits the behavioral
information to node 8 successfully, and nodes 3 and
6 become recovered. (d) At $t=3$, nodes 4 and 5 are activated
in the instantaneous structure $G_3$. The contagion process terminates
since all adopted nodes become recovered. }
\label{fig1}
\end{center}
\end{figure}

We generate a time-varying community network based on
the activity driven network model~\cite{Perra2012}.
To simplify analysis, we suppose a network with $N$ nodes (representing individuals), consisting of two communities $A$ and $B$ with equal sizes.
Initially, each node is assigned an equal activity potential $a$. The instantaneous network structure $G_t$ is generated as below:
At time step $t$, each node is activated with probability $a$. If a node $v$ is activated, it will generate
$m$ edges, each of which randomly connects to a
node in the same community with probability
$\mu$, called community strength, and connect to a node in the different
community with probability $1-\mu$~[see Fig.~\ref{fig1}]. Multiple edges and self-loops are not
allowed. In order to form
community structures, we set $\mu\in[0.5,1]$. Obviously, there
will be less edges between the communities with
the increase of $\mu$. For a small value of $\mu$, the
community structure is not obvious. Note that when
$\mu=0.5$, the probabilities of an edge connecting to the same
and the different communities are equal, and the
edges are connected completely randomly, thus the time-varying community structures disappear. When $\mu=1$, there is
no edges between communities, leading to two totally isolated
communities. At the end of time step $t$, we delete all the generated
edges. Repeating the above process generates a time-varying community network.

\subsection{Social contagion model}
\label{2.2}
We propose a non-Markovian social contagion model, called susceptible-adopted-recovered (SAR) model,
to describe behavior spreading on time-varying
community networks~\cite{wangwei2015a,Shu2016}. At a given time step, a
node can be in one of the three states: susceptible, adopted,
and recovered. In the susceptible state,
a node has not adopted the behavior and is willing to receive behavioral information from its neighbors who has adopted the behavior. In the adopted state, a
node who has adopted the behavior and is keen to spread
the behavioral information to its neighbors. In the recovered state, a node
will lose its interest to the behavior and no longer anticipate
the spreading process. Each node
holds a static equal adoption threshold $\pi$, which reflects the wills
of this node to adopt the behavior. Each node has
variable $\chi_i$ to count how many
pieces of behavioral information
it has received.

At the beginning, a proportion
$\rho_0$ of nodes are randomly chosen as seeds (initial adopters), while
the remaining nodes are susceptible. We use synchronous
updating method to update nodes' state~\cite{wangwei2015a}.
At each time step, we first generate
an instantaneous structure $G_t$ according to the method described in
Sec.~\ref{2.1}. Then, the behavior spreads on network $G_t$ as
follows. Every adopted node $v$ transmits the behavioral
information to each susceptible neighbor $u$ with probability
$\lambda$. If $u$ receives the information successfully, his
corresponding accumulated information counter $\chi_u$ will add one.
If $\chi_u$ reaches or exceeds the adoption threshold $\pi$, the node
$u$ becomes
adopted state. The dynamics of social contagion
is a non-Markovian stochastic process. For the case of $\pi=1$,
the model becomes memoryless,
thus we only discuss the situations when $\pi>1$. At the same
time step, the adopted nodes
become recovered with probability
$\gamma$. The contagion process terminates when all adopted
nodes become recovered. In this model, the probabilities $\lambda$ and $\gamma$ can be interpreted as transmission rate and
recover rate respectively, for they are expected to equal to the proportion of information successfully arrived and nodes turning into
recover state at each time step. An illustration of our social contagion model on time-varying community network is given in Fig.~\ref{fig1}.

\section{Theory} \label{sec:theory}
In this section, we develop a mean-field approximation theory
to quantitatively describe the non-Markovian social contagions
on time-varying community network. We denote the proportion of
susceptible nodes who have received $r$ pieces
of behavioral information in community $A$ and community
$B$ at time step $t$ as $S_A(r,t)$ and $S_B(r,t)$ (denominator is $N/2$), respectively.
We respectively use $\rho_A(t)$ and $\rho_B(t)$ to denote
the proportion of adopted nodes in communities $A$ and $B$, and $R_A(t)$ and $R_B(t)$ to denote the proportion of
recovered nodes at time step $t$. When $t\rightarrow\infty$,
all adopted nodes become recovered. We denote the final proportion
of nodes in the recovered state in communities $A$ and $B$
as $R_A(\infty)$ and $R_B(\infty)$, respectively. The final behavior
adoption proportion in the whole network is then $R(\infty)=[R_A(\infty)+R_B(\infty)]/2$, since communities $A$ and $B$ have the same size.

Due to the symmetry of the two communities, we only
introduce the theoretical analyses on community $A$ detailedly, and the results
on community $B$ can be derived by simply exchanging the index
$A$ and $B$. At time step $t$, a node $v_A$ forms its $k$ edges in the instantaneous structure $G_t$ in
two different ways: (i) edges generated by $v_A$ itself,
denoted as its out-going degree $k_o$; (ii) edges generated by
other active nodes in the network connecting to $v_A$, denoted as its in-coming
degree $k_i$. As a result, the degrees of node $v_A$ is $k=k_i
+k_o$. One can assume that the degrees of active nodes are equal,
and the degree of inactive nodes are also the same. According to the formation of the time-varying community networks described in Sec.~\ref{sec:model},
node $v_A$ generates $m$ edges
to connect to other nodes when it is active, thus $k_o=m$.
At the same
time, other active nodes in the network
generate edges and try to connect to $v_A$. For there are expected $am(N-1)$
edges to be remained on $G_t$, node $v_A$ will get
$k_i=(N-1)am/N\approx ma$ connections since the communities $A$
and $B$ are symmetric. Thus, we obtain the expected degree of active nodes
as $k=m+ma$~\cite{Lixiang2014}.
When $v_A$ is inactive, $k_o=0$ while $k_i$ remains the same,
thus $k=k_i=ma$. For active nodes, each of its edges connects to
a node in the same community
with probability $\mu$ and connects to the different community with probability $1-\mu$.
The probability that node $v_A$ connecting to $n$
nodes in community $A$ when it is active
can thus be written as
\begin{equation} \label{1}
\omega^{AA}_{Active}(n)=\binom{m+ma}{n}\mu^n(1-\mu)^{m+ma-n}.
\end{equation}
Similarly, the probability that node
$v_A$ connecting to $n$ nodes in community $B$ is given by
\begin{equation}
\omega^{AB}_{Active}(n)=\binom{m+ma}{n}(1-\mu)^n\mu^{m+ma-n}.
\end{equation}
If node $v_A$ is inactive, the probability
that $v_A$ has $n$ edges connecting to nodes in community $A$ or $B$ are
\begin{equation} \label{1a}
\omega^{AA}_{Inactive}(n)=\binom{ma}{n}\mu^n(1-\mu)^{ma-n}
\end{equation}
and
\begin{equation}
\omega^{AB}_{Inactive}(n)=\binom{ma}{n}(1-\mu)^n\mu^{ma-n},
\end{equation}
respectively.

On the instantaneous structure $G_t$, the probability that a node
$v_A$ in community $A$ with degree $k=k_A$ has
$x_A$ adopted neighbors in community $A$ is
\begin{equation} \label{2}
\xi_{AA}(k_{AA},x_A,t)=\binom{k_{AA}}{x_A}
[\rho_A(t)]^{x_A}[1-\rho_A(t)]^{k_{AA}-x_A},
\end{equation}
where $k_{AA}$ denotes the number of neighbors of node $v_A$ in community $A$. Similarly, the probability that
$v_A$ has $x_B$ adopted neighbors in community
$B$ can be written as
\begin{equation} \label{3}
\xi_{AB}(k_{AB},x_B,t)=\binom{k_{AB}}{x_B}
[\rho_B(t)]^{x_B}[1-\rho_B(t)]^{k_{AB}-x_B},
\end{equation}
where $k_{AB}$ is the number of neighbors of
$v_A$ in community $B$ and $k_{AB}=k_A-k_{AA}$.

We separately consider the situations that $v_A$ is active
or inactive at time step $t$. For the former situation,
combining Eqs.~(\ref{1}) and (\ref{2})-(\ref{3}), the probability
that the active $v_A$ connects to $n$ adopted nodes is
\begin{equation} \label{4}
\begin{split}
\theta^A_{Active}(n,t)&=\sum^{m+ma}_{i=0}\omega^{AA}_{Active}(i)
\sum^{{\rm min}(n,i)}_{j=0}[\xi_{AA}(i,j,t)\\
&\times\xi_{AB}(m+ma-i,n-j,t)],
\end{split}
\end{equation}
where we use ${\rm min}(x,y)$, meaning the minimum value of $x$ and $y$, to avoid the situations that $j$ exceeds $i$. When node
$v_A$ is inactive at time step $t$, the probability that $v_A$ connects
to $n$ adopted nodes can be obtained by combining Eqs.~(\ref{1a})
and (\ref{2})-(\ref{3}),
\begin{equation} \label{5}
\begin{split}
\theta^A_{Inactive}(n,t)&=\sum^{ma}_{i=0}\omega^{AA}_{Inactive}(i)
\sum^{{\rm min}(n,i)}_{j=0}[\xi_{AA}(i,j,t)\\
&\times\xi_{AB}(ma-i,n-j,t)].
\end{split}
\end{equation}
Summarize the two situations and combining Eqs.~(\ref{4})-(\ref{5}),
the probability that $v_A$ connects to $n$ adopted
individuals on $G_t$ is given by
\begin{equation} \label{6}
\theta_{A}(n,t)=a\theta^A_{Active}(n,t)+(1-a)\theta^A_{Inactive}(n,t).
\end{equation}

Then we focus on the time evolution of the
density of nodes in each state.
According to the social contagion model, when a susceptible node
$v_A$ has $n$ adopted neighbors at time step $t$, the probability
that it receives at least one piece of
behavioral information from its neighbors is
\begin{equation} \label{7}
\psi_A(t)=\sum^{m+ma}_{n=1}\theta_{A}(n,t)[1-(1-\lambda)^n].
\end{equation}
The probability that $v_A$ receives
$i\geq1$ pieces of behavioral information can be expressed as
\begin{equation} \label{8}
\phi_A(i,t)=\sum^{m+ma}_{n=i}\theta_{A}(n,t)\binom{n}{i}
\lambda^i(1-\lambda)^{n-i}.
\end{equation}
Obviously, Eq.~(\ref{7}) can be derived
by Eq.~(\ref{8}) as
\begin{equation} \label{9}
\psi_A(t)=\sum^{m+ma}_{i=1}\phi_A(i,t).
\end{equation}

Then the time evolution of the contagion process can be described
by a developed mean-field method. For those nodes who
have not received any behavioral information at time step $t$, denoted
as $S_A(0,t)$, they change into other states when receiving at
least one piece of behavioral information, yielding
\begin{equation} \label{10}
\frac{dS_A(0,t)}{dt}=-S_A(0,t)\psi_A(t).
\end{equation}
When $1\leq r<\pi$, the increase of $S_A(r,t)$ comes
from these nodes who have only received less than
$r$ pieces of behavioral information, that is $S_A(q,t)$
($0\leq q<r$), change into $S_A(r,t)$ after receiving $r-q$
pieces of behavioral information, with probability
$\sum^{r-1}_{q=0}S_A(q,t)\phi_A(r-q,t)$. At the same time,
$S_A(r,t)$ decreases after those nodes receive at least one information
and then turns to other states, with the probability $S_A(r,t)
\psi_A(t)$. Thus, the evolution equation of $S_A(r,t)$ can be
written as

\begin{equation} \label{11}
\frac{dS_A(r,t)}{dt}=\sum^{r-1}_{q=0}S_A(q,t)\phi_A(r-q,t)-S_A(r,t)\psi_A(t).
\end{equation}

Similarly, the increase of adopted nodes results
from the state change of susceptible nodes who have
received information being equal or over the threshold $\pi$,
with probability $\sum^{\pi-1}_{q=0}S_A(q,t)[\psi_A(t)-
\sum^{\pi-1-q}_{i=1}\phi_A(i,t)]$, and the decrease owes
to the recovering of themselves, with probability $\gamma
\rho_A(t)$. Thus the evolution
of the densities of adopted and recovered nodes can be written as
\begin{equation} \label{12}
\frac{d\rho_A(t)}{dt}=\sum^{\pi-1}_{q=0}S_A(q,t)[\psi_A(t)-
\sum^{\pi-1-q}_{i=1}\phi_A(i,t)]-\gamma\rho_A(t)
\end{equation}
and
\begin{equation} \label{13}
\frac{dR_A(t)}{dt}=\gamma\rho_A(t),
\end{equation}
respectively.

Now, Eqs.~(\ref{10})-(\ref{13}) form a complete description
of the social contagion process,
allowing us to compute the proportion of nodes in any state in community $A$ at any time step.
By transferring our knowledge to community $B$ i.e., exchanging the positions of
index $A$ and $B$, the time evolutions in community
$B$ and in the whole network are also available. When all
adopted nodes become recovered, we count the final
behavior adoption proportion $R(\infty)=[R_A(\infty)+R_B(\infty)]/2$.

The outbreak threshold of social contagion $\lambda_c$
is a crucial parameter. When the
information transmission rate $\lambda$ is greater
than $\lambda_c$, a finite fraction of
nodes adopt the behavior.
When $\lambda\leq\lambda_c$, there is only a vanishingly small
fraction of nodes adopting the behavior. Initially, there are a few nodes in the
adopted state, thus $\rho_A(0)\rightarrow0$, $\rho_B(0)
\rightarrow0$, $R_A(0)\rightarrow0$ and $R_B(0)\rightarrow0$.
Previous studies indicate that the behavior can
outbreak over the network, if and only if the proportion
of adopted individuals can exponentially grow at initial time~\cite{WangWeiSR,
NewmanBook}. Thus, one expects to
obtain $\lambda_c$ by stability analysis method.
Unfortunately, this
method is useless to our model because of the memory effect.
On the one hand, a vanishingly
small fraction of initial adopters can not lead to the
quick growth of behavior at initial time in our model, for the susceptible nodes cannot immediately accumulate the information memory to reach or exceed the adoption threshold $\pi$ when the initial adopters are very rare.
On the other hand, the appearance of nonlinearity in the
system makes the linearization method near the stability
point ineffective~\cite{Radicchi2015}. Therefore, the outbreak
threshold can not be obtained by the existing method. To get the
outbreak threshold, further researches are needed.

\section{Simulation Results}\label{sec:sim}
Based on the proposed model, we performed extensive simulations to investigate the social contagions on time-varying community networks. In simulations, the size of network, recover
probability and adoption threshold are set to be $N=10,000$, $\gamma=0.1$
and $\pi=3$, respectively. At the beginning, a proportion $\rho_0=0.06$ of nodes in Community $A$ are randomly
chosen as seeds, while the
remaining nodes are susceptible.
The simulation results of the final adoption proportion $R_A(\infty)$, $R_B(\infty)$ and $R(\infty)$ are obtained by averaging the results over $2000$ independent realizations. The theoretical values
of $R_A(\infty)$, $R_B(\infty)$ and $R(\infty)$ are
given by solving Eqs.~(\ref{10})-(\ref{13}). We separately discuss the effects of community structure
 and the time-varying structure on the social contagions.

\subsection{Effects of community structure}

\begin{figure}
\begin{center}
\epsfig{file=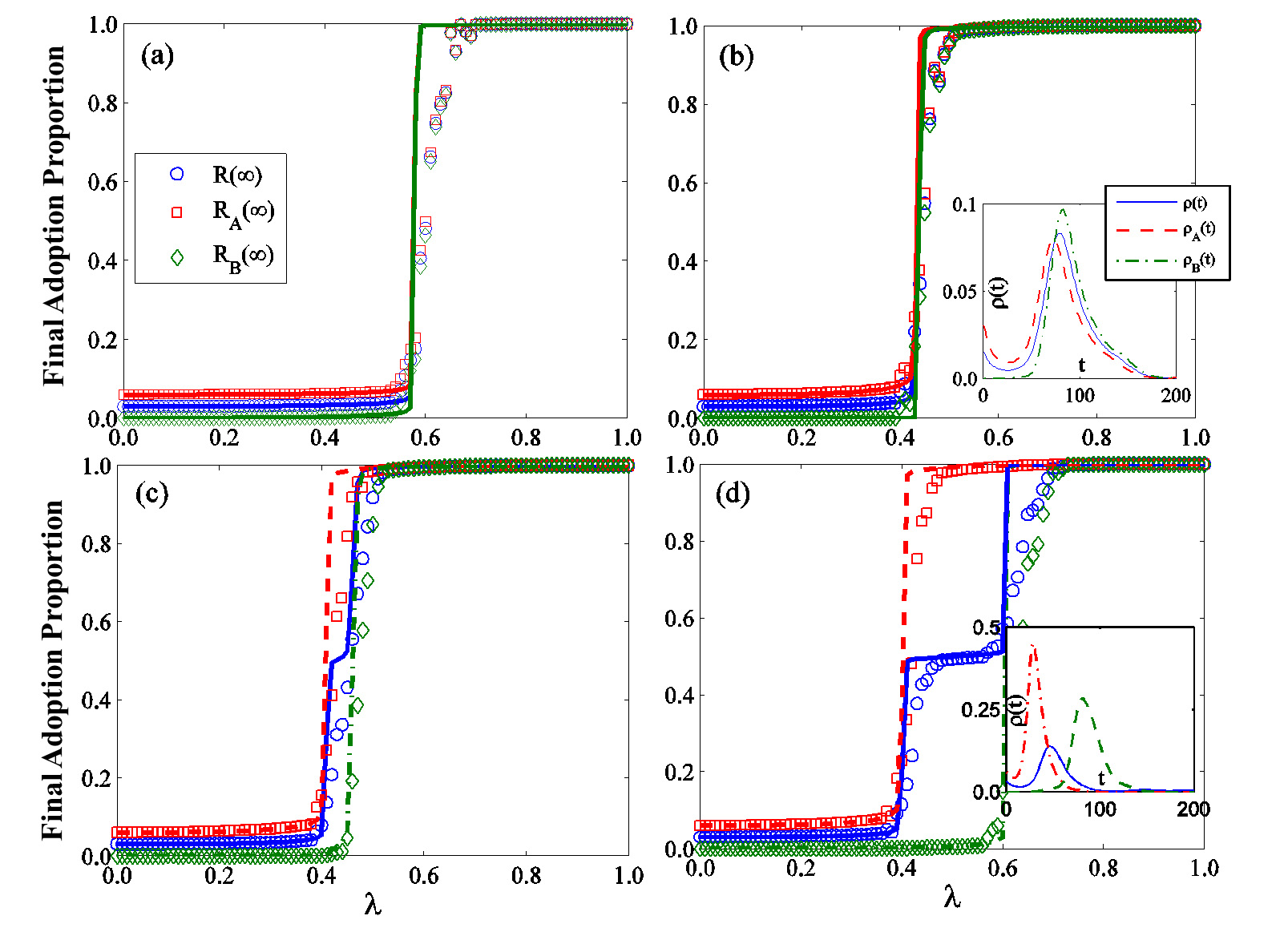,width=1\linewidth}
\caption{(Color online) The final behavior adoption proportion $R_A(\infty)$,
$R_B(\infty)$ and $R(\infty)$ versus information transmission
probability $\lambda$ under different community strengths.
(a) $\mu=0.5$, (b) $\mu=0.9$,
(c) $\mu=0.95$, and (d) $\mu=0.97$. The solid line (circles),
dotted line (squares) and dotted line (diamonds) represent the theoretical
predictions (simulation results) of $R_A(\infty)$, $R_B(\infty)$ and $R(\infty)$.
The insets of (b) and (d) show simulation
results of
$\rho_A(t)$, $\rho_B(t)$ and $\rho(t)$ versus $t$. Other
parameters are set to be $N=10,000$, $\rho_0=0.03$, $a=0.2$,
$m=5$, $\gamma=0.1$ and $\pi=3$, respectively. }
\label{fig2}
\end{center}
\end{figure}

We firstly study the growths of $R_A(\infty)$, $R_B(\infty)$
and $R(\infty)$ versus $\lambda$ under different
$\mu$ in Fig.~\ref{fig2}, which
show different growth patterns. For relatively small
values $\mu=0.5$ and $0.9$,
nodes in community $A$ and community $B$ adopt the
behavior at almost the same time [see Figs.~\ref{fig2}(a)-(b)].
That is because the community structure is not obvious
when $\mu$ is relatively small, and the adopted nodes
are able to transmit the behavioral information to
nodes in the whole network. For relatively large
values of $\mu=0.95$ and $0.97$,
the behavior adoption exhibits
a hierarchical feature: nodes in community
$A$ first adopt the behavior, and then nodes in community $B$
adopt the behavior with the increase of $\lambda$
[Figs.~\ref{fig2}(c)-(d)]. When $\mu$ is relatively large,
nodes tend to transmit information to those nodes
in the same community, which adds difficulty to transmit
the information to community $B$. The insets of Figs.~\ref{fig2}
(b) and (d) show the corresponding growth patterns of $\rho_A(\infty)$,
$\rho_B(\infty)$ and $\rho(\infty)$ versus time $t$, which
confirms the hierarchical feature shown in the behavior adoption process.
The theoretical predictions
agree well with the
simulation results, giving a quantitative description of
the above phenomena.
The deviations between the theoretical predictions and the
simulation results are caused by the dynamical correlations among
the states of the neighbors and finite-size network
effects~\cite{Reche2011,Cui2014}.

\begin{figure}
\begin{center}
\epsfig{file=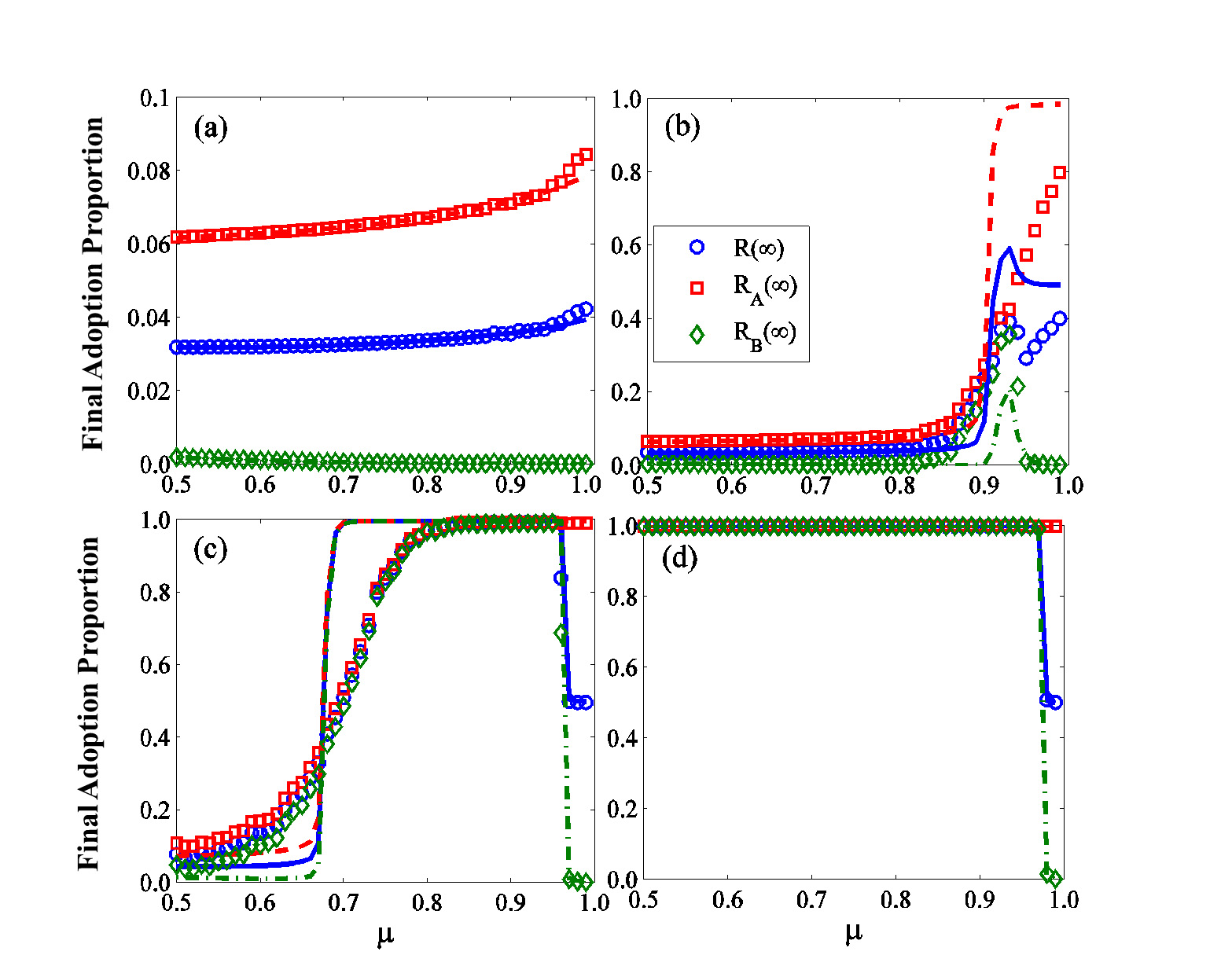,width=1\linewidth}
\caption{(Color online) The final behavior adoption proportion $R_A(\infty)$,
$R_B(\infty)$ and $R(\infty)$ versus community strength $\mu$
under different information transmission rates
(a) $\lambda=0.35$, (b) $\lambda=0.4$3, (c) $\lambda=0.55$ and
(d) $\lambda=0.7$. The
solid line (circles), dotted line (squares) and dotted line (diamonds)
represent the theoretical values (simulation values) of $R_A(\infty)$,
$R_B(\infty)$ and $R(\infty)$, respectively.
Other
parameters are set to be $N=10,000$, $\rho_0=0.03$, $a=0.2$,
$m=5$, $\gamma=0.1$ and $\pi=3$, respectively.}
\label{fig3}
\end{center}
\end{figure}

Figure \ref{fig3} exhibits the growths of $R_A(\infty)$,
$R_B(\infty)$ and $R(\infty)$ versus $\mu$ under different
$\lambda$. Three different growth patterns can be observed.
For small values of $\lambda$ in Fig.~\ref{fig3}(a),
$R(\infty)$, $R_A(\infty)$ and $R_B(\infty)$ monotonically
increase with growing $\mu$. With the increase of $\lambda$,
shown in Figs.~\ref{fig3} (b)-(c), $R_A(\infty)$ increases
with $\mu$ monotonically, while $R_B(\infty)$ and $R(\infty)$
first increase and then decrease, which indicates the existence of optimal
community strength promoting the behavior adoption.
The optimal contagion phenomena can be explained as below:
There are more edges in the community for a larger $\mu$, which promotes the behavior spreading on community
$A$. Meanwhile, the amount of bridge edges between communities decreases
with growing $\mu$. If $\mu$ is large enough, the global behavior
adoption will be inhibited, which leads to the decrease of
$R_B(\infty)$ and $R(\infty)$. When $\lambda$ is very
large, nodes in both communities
adopt the behavior easily [see Fig.~\ref{fig3}(d)]. For any given $\mu$, the $R_A(\infty)$ can always reach a remarkable value.
Only when $\lambda$ is great enough, the two communities
tend to be isolated and the global behavior adoption is suppressed,
thus $R_B(\infty)$ and $R(\infty)$ begin to decrease.

\begin{figure}
\begin{center}
\epsfig{file=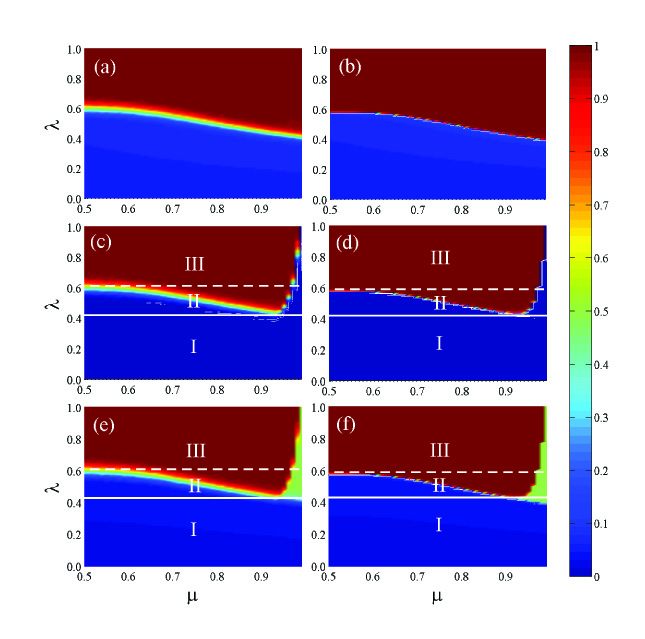,width=1\linewidth}
\caption{(Color online) The final behavior adoption proportion $R_A(\infty)$,
$R_B(\infty)$ and $R(\infty)$ versus community strength $\mu$
and information transmission rate $\lambda$.
Color-coded values show simulation results in (a) $R_A(\infty)$
,(c) $R_B(\infty)$ and (e) $R(\infty)$, and theoretical predictions
in (b) $R_A(\infty)$, (d) $R_B(\infty)$ and (f) $R(\infty)$. Other
parameters are set to be $N=10,000$, $\rho_0=0.03$, $a=0.2$,
$m=5$, $\gamma=0.1$ and $\pi=3$, respectively.}
\label{fig4}
\end{center}
\end{figure}

We show the effects of $\lambda$ and $\mu$ on
$R_A(\infty)$, $R_B(\infty)$ and $R(\infty)$ in
Fig.~\ref{fig4}. According to the growth patterns of
$R(\infty)$, $R_A(\infty)$ and $R_B(\infty)$ versus $\mu$ in
Figs.~\ref{fig4}(c)-(f), $\mu$-$\lambda$ plane can be divided
into three regions: (I) monotonically increasing region,
(II) non-monotonically changing region and (III) monotonically
decreasing region. As $R_A(\infty)$ increases
monotonically with $\mu$, Figs.\ref{fig4}(a)-(b) only exist region I.
Due to the effect of time-varying community structures,
Figs.~\ref{fig4}(c)-(f) exhibits three different regions, which
means that there exists an optimal community strength in a
certain range of $\lambda$, making the values of $R_B(\infty)$ and $R(\infty)$
reach the maximum values. The theoretical results in
Fig.~\ref{fig4}(b),(d),(f) can well predict the simulation
results in Figs.~\ref{fig4}(a),(c),(e).

\subsection{Effects of time-varying structure}

\begin{figure}
\begin{center}
\epsfig{file=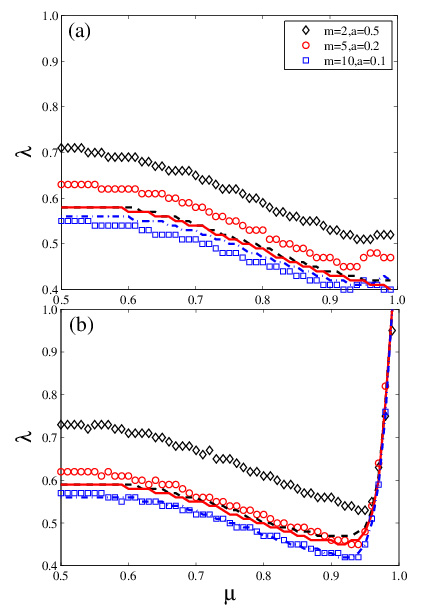,width=1\linewidth}
\caption{(Color online) The behavioral
information transmission rate $\lambda$ versus community strength $\mu$
for a given behavior adoption proportion (a) $R(\infty)=0.4$
and (b) $R(\infty)=0.7$. The
dotted line (diamonds), solid (circles) and dotted line (squares)
denote the theoretical values (simulated values) of $m=2$, $a=0.5$,
$m=5$, $a=0.2$ and $m=10$, a=0.1, respectively Other
parameters are set to be $N=10,000$, $\rho_0=0.03$, $a=0.2$,
$m=5$, $\gamma=0.1$ and $\pi=3$, respectively.}
\label{fig5}
\end{center}
\end{figure}

In Fig.~\ref{fig5}, we investigate the effect of
time-varying structure on social contagions. According
to the description of the time-varying
community structures, the average degree of
$G_t$ is $\langle k\rangle=2ma$ at time step $t$, which allows
us to compare the relative importance of time-varying structure
parameters $m$ and $a$ on the social contagions. We keep
the rest of parameters the same as Fig.~\ref{fig2} and
fix the average degree $\langle k\rangle=2$, then adjust
the values of $m$ and $a$.
For a given $R(\infty)$ and $\mu$, we record the corresponding
values of $\lambda$,
i.e., getting the contours
of different $R(\infty)$ in the $\lambda$-$\mu$ plane.
If the importance of $m$ and $a$ are equal, their
influences on social contagions will counteract each other, and the simulation results will remain almost the same. However, we find that the
information transmission rate $\lambda$ needed to
reach the specified $R(\infty)$ decreases
with the increase of $m/a$, which implies that, compared
to adjusting the the value of $a$, adjusting $m$ is
more beneficial to social contagions.
We can understand the
phenomenon in the following way: Increasing the value of $m/a$ means decreasing the number of active nodes and increasing the average degree of active nodes, which leads to emerge of active nodes with high degree. When active nodes have high degree, they will have high probability to touch enough adopted nodes and become adopted at one time step, thus these contacts are effective. On the contrary, small $m/a$ will result in low degree of active nodes in instantaneous structure. These active nodes will not receive enough information at one time step, and wait for another round of activating, which is not so effective. Though active nodes existing at one step are few because of small $a$, high degree situation can be more efficiently, and eventually reach the assigned $R(\infty)$ more quickly.
For other average degree, such as $\langle k\rangle=0.2,1,3$, the same phenomena can be observed.
Our theoretical method also displays the same phonomania about the effects of $m/a$ in Fig.~\ref{fig5}.

\section{Discussion} \label{sec:dis}
In this paper, we studied the effects of time-varying community structures on social contagions.
We first proposed a non-Markovian
social contagion model on time-varying community
network, and then develop a mean-field theory to
quantitatively describe the proposed model. Through theoretical
analyses and extensive numerical simulations, we found that
behavior adoption exhibits a hierarchical
feature. The behavior first spreads in
one of the communities, and then outbreaks in the other. Moreover,
under different behavioral information
transmission rates, the final behavior
adoption proportion in the whole network versus the community
strength can show one of the different patterns, such as, a monotone increasing pattern, a non-monotonic changing pattern,
and a monotone decreasing pattern. In non-monotonic changing pattern, we found an optimal community strength under which
the final behavior adoption proportion reaches its maximum value. Finally, we discovered that for
a given average degree, increasing the number of edges generated
by active nodes is more beneficial to the social contagions than increasing the average
activity potential. Our proposed theory
predicted the phenomena on social contagion well.

We qualitatively and quantitatively studied how time-varying
community structures affect the social contagions.
First, we described timeliness of the edges by using the
time-varying network model, compensating the lack of static network
research methods. In addition, the proposed non-Markovian
social contagion model described social contagion process
on time-varying community network more accurately than Markovian models.
Furthermore, our developed theory predicted qualitatively
the occurrence of various phenomena in simulations. In
conclusion, this work helps us in better understanding,
predicting and controlling the social contagions
on social networks. The effects of social
contagions on epidemic spreading and the relationship between time-varying networks and multilayer networks are worthy of future study~\cite{Ruan2012,YangHX2015,Arenas2013,Arenas2014}.

\acknowledgments

This work was supported by the National Natural Science Foundation of China under Grants Nos. 11105025, 11575041, 61433014, and 61473001,
and the Fundamental Research Funds for the Central Universities (Grant No. ZYGX2015J153),
and the Scientific Research Starting Program of Southwest Petroleum University (Grant No. 2014QHZ024).

%\bibliographystyle{apsrev4-1}
%\bibliography{Social_Contagion}

\end{document}